\documentclass{ws-procs961x669-vaso} 
\usepackage{hyperref}
\usepackage{xspace}  
\usepackage{slashed}

\DeclareGraphicsExtensions{.pdf,.png,.jpg} 
\graphicspath{{./}{./figs/}}
\newcommand{\ifb}{\mbox{fb$^{-1}$}\xspace}
\newcommand{\mev}{\mbox{MeV}\xspace}
\newcommand{\gev}{\mbox{GeV}\xspace}
\newcommand{\tev}{\mbox{TeV}\xspace}
\newcommand{\none}{\ensuremath{\tilde{\chi}_1^0}\xspace}

\begin{document}

\mark{{V.~A.~Mitsou}{LHC experiments for long-lived particles of the dark sector}}

\vspace*{-1.4cm}
\begin{flushright}
IFIC/21-44
\end{flushright}

\title{LHC experiments for long-lived particles of the dark sector}

\author{Vasiliki~A.~Mitsou}

\address{Instituto de F\'isica Corpuscular (IFIC), CSIC -- Universitat de Val\`encia, \\
C/ Catedr\'atico Jos\'e Beltr\'an 2, E-46980 Paterna (Valencia), Spain\\
E-mail: vasiliki.mitsou@ific.uv.es\\
webific.ific.uv.es/web}

\begin{abstract}
Dark matter scenarios are being tested at the LHC in the general-purpose experiments through promptly decaying states. In parallel, new dedicated detectors have been proposed for the LHC to probe dark matter portal theories predicting long-lived particles that decay away from the interaction point: MoEDAL-MAPP, MoEDAL-MALL, FASER, SND@LHC, CODEX-b, MATHUSLA, AL3X, ANUBIS, FACET, milliQan, FORMOSA. In addition, the SHiP beam-dump experiment is planned to operate with the SPS beam to extend the discovery reach for such particles. The detector design and expected physics sensitivity of these experiments is presented with emphasis on scenarios explaining the nature of dark matter.
\end{abstract}


\bodymatter

\section{Introduction}\label{sc:intro}

In collider physics, and in particular at the Large Hadron Collider (LHC),~\cite{Evans:2008zzb} there is a growing experimental interest in long-lived particles (LLPs),~\cite{Alimena:2019zri,Lee:2018pag} which travel a macroscopic distance before either decaying within the detector or giving rise to anomalous ionisation. In theory, long lifetimes may be due to a symmetry (leading to stable particles), narrow mass splittings, small couplings, or a heavy mediator. In this arena, a class of LLPs called Feebly interacting Particles (FIPs)~\cite{Agrawal:2021dbo} is characterised by indirect interactions of FIPs with Standard Model (SM) particles through low-dimensional operators. These interactions, commonly referred to as \emph{portals}, may predict dark photons (vector portal), a light dark Higgs boson (scalar portal), axion-like particles (pseudoscalar portal), or heavy neutral leptons (fermion portal). 

Scenarios beyond-the-SM (BSM) that introduce a hidden sector in addition to the visible SM sector are required to explain a number of observed phenomena in particle physics, astrophysics and cosmology such as the non-zero neutrino masses, the dark matter (DM), the baryon asymmetry of the Universe and the cosmological inflation. The mystery of DM, in particular, is quite intriguing with around 25\% of our Universe invisible (dark) that only interacts through gravity and remains unaccounted for in the SM. This review focuses on the possibility to explore DM with LLPs, as a complementary way to searches performed by the main experiments. ATLAS~\cite{ATLAS:2008xda} and CMS~\cite{CMS:2008xjf} look for DM in mono-$X$ final states, in associated production and in resonances via mediators~\cite{Mitsou:2013rwa}, while LHCb~\cite{Alves:2008zz} has good prospects for probing DM-portal LLPs after its Phase-I upgrade.~\cite{Borsato:2021aum} 

This review is organised as follows. After a brief introduction in \sref{sc:intro}, the specially designed LHC experiments to tackle LLPs are concisely presented in \sref{sc:experiments}. The dark-scalar and dark-vector portals, proposing neutral LLPs that decay in the detector volumes to known particles, are reviewed in \sref{sc:higgs} and \sref{sc:dp}, respectively. Dark photons at the limit of zero mass, leading to fractionally charged particles, are discussed in \sref{sc:mcp}. In \sref{sc:other}, other BSM scenarios pertinent to LLPs are highlighted. Finally, possible measurements of new particle properties are outlined in \sref{sc:discrim}, before summarising and giving an outlook in \sref{sc:summary}.

\section{Dedicated LLP Experiments at the LHC}\label{sc:experiments}

There is a constantly growing list of experiments planned for the LHC optimised for the detection of feebly interacting particles. Their projected added value stems mostly from their different operation and geometry parameters: angle w.r.t.\ the beam axis, the detector volume and distance from IP --- probing different ranges of lifetime, couplings to SM and boost (mass) ---, their time scale and the detector design.

Besides the detectors highlighted below, there are other, operating or future, non-LHC experiments with sensitivity to DM and other portal models. Such cases include BaBar,~\cite{BaBar:2017tiz} existing CERN beam-dump experiments, like the NA48/2,~\cite{NA482:2015wmo} NA62~\cite{NA62:2020xlg} and NA64,~\cite{Banerjee:2019pds} the Fermilab SeaQuest~\cite{Berlin:2018pwi} and the reactor experiment SoLid.~\cite{Roy:2021mke} Proposed experiments include the Search for Hidden Particles (SHiP)~\cite{SHiP:2015vad} at the CERN Beam Dump Facility, SHADOWS~\cite{Baldini:2021hfw} at the CERN North Area, LUXE-NPOD~\cite{Bai:2021dgm} at DESY, HECATE~\cite{Chrzaszcz:2020emg} for the FCC-ee or CEPC, GAZELLE~\cite{Dreyer:2021aqd} in Bell~II at SuperKEKB, SUBMET~\cite{Choi:2020mbk} at J-PARC and FerMINI~\cite{Kelly:2018brz} at Fermilab, among others. 

\subsection{MAPP -- MoEDAL Apparatus for Penetrating Particles}\label{sc:mapp}

The MoEDAL (Monopole and Exotics Detector at the LHC)~\cite{Pinfold:2009oia} experiment is mainly dedicated to searches for manifestations of new physics through highly ionising particles in a manner complementary\cite{DeRoeck:2011aa} to ATLAS and CMS. It is the first dedicated \emph{search} LHC experiment. The principal motivation for MoEDAL is the quest for magnetic monopoles~\cite{Mavromatos:2020gwk} and dyons, as well as for any massive, stable or long-lived, slow-moving particle~\cite{Fairbairn:2006gg} with single or multiple electric charge arising in various extensions of the SM.~\cite{Acharya:2014nyr} The MoEDAL detector~\cite{Mitsou:2020hmt} is deployed around the region at interaction point~8 (IP8) of the LHC in the LHCb~\cite{Alves:2008zz} vertex locator cavern. 

It is a unique and, to a large extend, passive detector based on three different detection techniques:  nuclear track detectors (NTDs), magnetic monopole trackers (MMTs) and TimePix pixel devices. MoEDAL has pioneered the quest for magnetic charges, by being the only contender in the high-charge regime~\cite{MoEDAL:2016jlb,Acharya:2016ukt,Acharya:2017cio,Acharya:2019vtb} and by being first in constraining dyons~\cite{MoEDAL:2020pyb} in colliders and in investigating the Schwinger thermal production of monopoles.~\cite{Acharya:2021ckc} More analyses, also on electric charges,~\cite{Acharya:2021yzs} are ongoing.~\cite{Mitsou:2021vhf}

\begin{figure}[ht]
\centering
\begin{minipage}[b]{0.57\linewidth}
  \includegraphics[width=\textwidth]{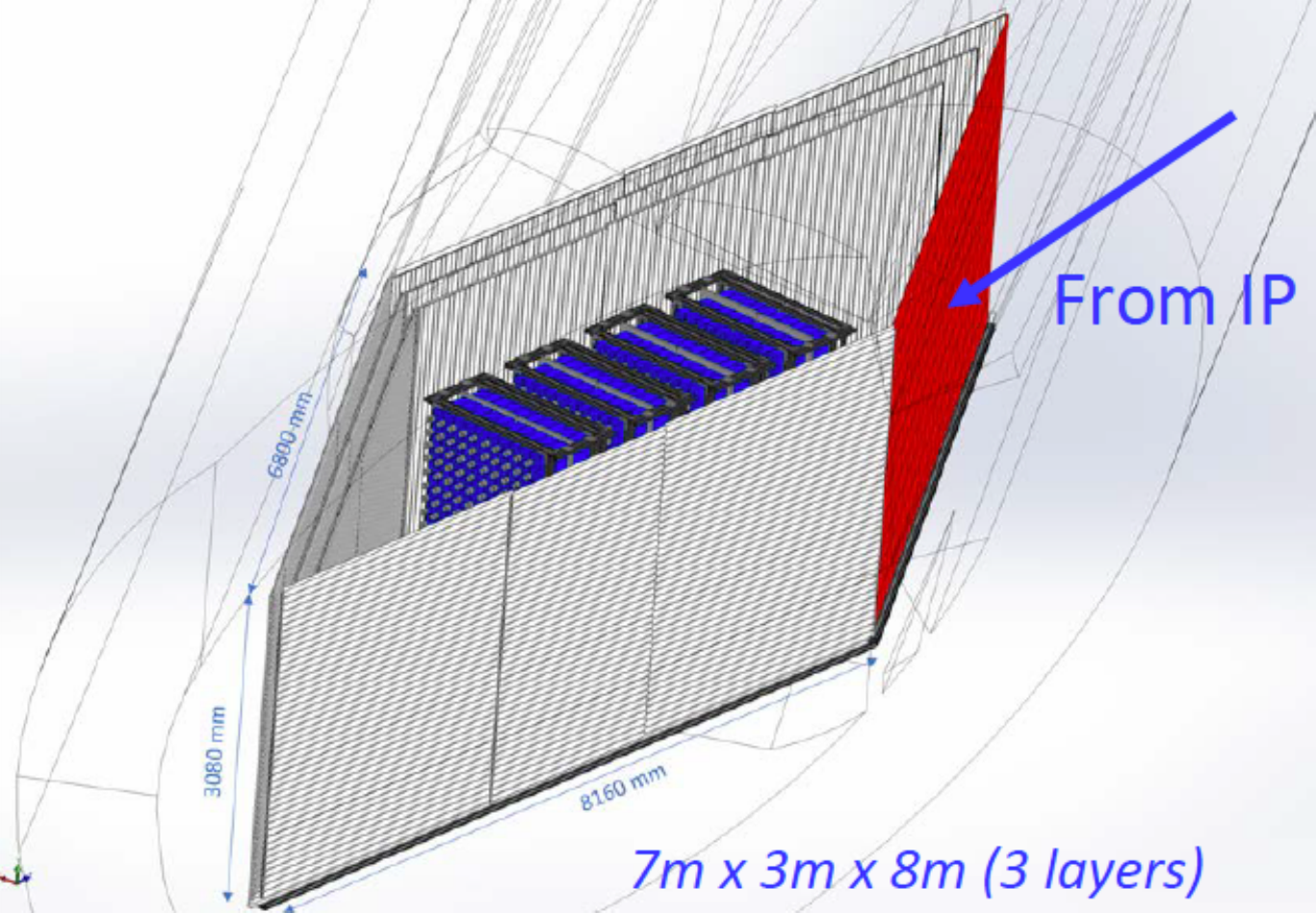}
\end{minipage}\hspace{0.03\linewidth}
\begin{minipage}[b]{0.39\linewidth}
  \caption{Diagram of the MAPP-1 detector components: mQP at the centre and three layers of LLP around it.} \label{fg:mapp}
\end{minipage}   
\end{figure}

The last few years, MoEDAL proposes to deploy MAPP~\cite{Pinfold:2019zwp,Pinfold:2019nqj} in a gallery near IP8 shielded by an overburden of approximately $100~{\rm m}$ of limestone from cosmic rays to extent its reach to the \emph{low} ionisation regime. It is envisaged that the first-stage detector, MAPP-1 will be installed during LHC Long Shutdown~2 (LS2) and Run~3.~\cite{Mitsou:2021vhf} The purpose of the innermost detector, the MAPP-mQP, shown in \fref{fg:mapp}, is to search for particles with fractional charge as small as $0.001e$, the so called \emph{millicharged particles (mCP)}, using plastic scintillation bars. A prototype of the mQP detector (10\% of the original system) is already in place since 2017 and the data analysis is ongoing.~\cite{Staelens:2019gzt} The Phase-1 MAPP-mQP is going to be deployed in the UA82 gallery in a distance $100~{\rm m}$ from IP8.

Another part of the detector, the MAPP-LLP, is deployed as three nested boxes of scintillator hodoscope detectors, in a ``Russian doll'' configuration, following as far as possible the contours of the cavern as depicted in \fref{fg:mapp}. It is designed to be sensitive to long-lived neutral particles from new physics scenarios via their interaction or decay in flight in a decay zone of size approximately $5~{\rm m} \text{ (wide)} \times 10~{\rm m} \text{ (deep)} \times 3~{\rm m} \text{ (high)}$. The MAPP detector can be deployed in a number of positions in the forward direction, at a distance of $\mathcal{O}(100~{\rm m})$ from IP8. An upgrade plan for the MAPP-1 detector is envisaged for the High-Luminosity LHC (HL-LHC),~\cite{Apollinari:2015wtw} called MAPP-2, with considerably larger volume than MAPP-1. 

Furthermore, the MoEDAL Apparatus for very Long Lived particles (MALL)~\cite{Pinfold:2019zwp} is intended to push the search for decays of new \emph{electrically charged}, massive and \emph{extremely} long-lived particles, with lifetimes well in excess of a year, by monitoring the exposed MMTs for decay products of trapped BSM particles.~\cite{Pinfold:2019zwp,Pinfold:2019nqj} Quite recently, an independent proposal has been presented to install such absorber volumes near the CMS IP.~\cite{Kieseler:2021esu} 

\subsection{FASER -- ForwArd Search ExpeRiment}\label{sc:faser}

FASER~\cite{Feng:2017uoz} is an approved small and inexpensive experiment designed to search for new particles produced in decays of light mesons copiously present at zero angle at the LHC in Run-3 and beyond. Such particles may be produced in large numbers and travel for hundreds of meters without interacting, and then decay to SM particles. To search for such events, FASER will be located $480~{\rm m}$ downstream of the ATLAS~\cite{ATLAS:2008xda} IP in the unused service tunnel TI12 (cf.\ \fref{fg:formosa}). It is planned to be constructed and installed in LS2 and collect data during Run~3.~\cite{FASER:2019aik} 

\begin{figure}[ht]
\centering
  \includegraphics[width=0.9\textwidth]{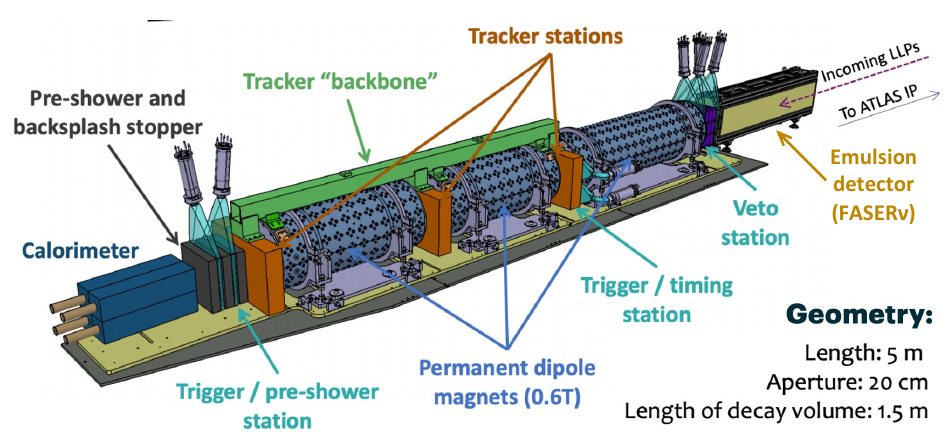}
  \caption{Annotated schematic view of the FASER detector components.} \label{fg:faser} 
\end{figure}

An overview of the different detector components is shown in \fref{fg:faser}.~\cite{Queitsch-Maitland:2021lmo} The magnets are 0.55~T permanent dipole magnets based on the Halbach array design with a radius of $10~{\rm cm}$. There are three scintillator stations for timing, trigger and vetoing incoming charged particles. The electromagnetic calorimeter consists of four spare outer ECAL modules from LHCb. The tracker consists of three tracking stations, each containing three layers instrumented with spare ATLAS semiconductor-tracker barrel modules.

Moreover, FASER$\nu$\cite{FASER:2019dxq} is designed to directly detect collider neutrinos and study their cross sections at \tev energies. In 2018, a pilot detector employing emulsion films was installed in the far-forward region of ATLAS, also visible in \fref{fg:faser}, and collected 12.2~\ifb of $pp$ collision data at a $\sqrt{s}=13~\tev$. The first candidate vertices consistent with neutrino interactions at the LHC were observed, an example of which is presented in \fref{fg:fasernu}, with a measured $2.7\sigma$ excess of neutrino-like signal above muon-induced backgrounds.~\cite{FASER:2021mtu}

\begin{figure}[ht]
\centering
  \includegraphics[height=0.19\textwidth]{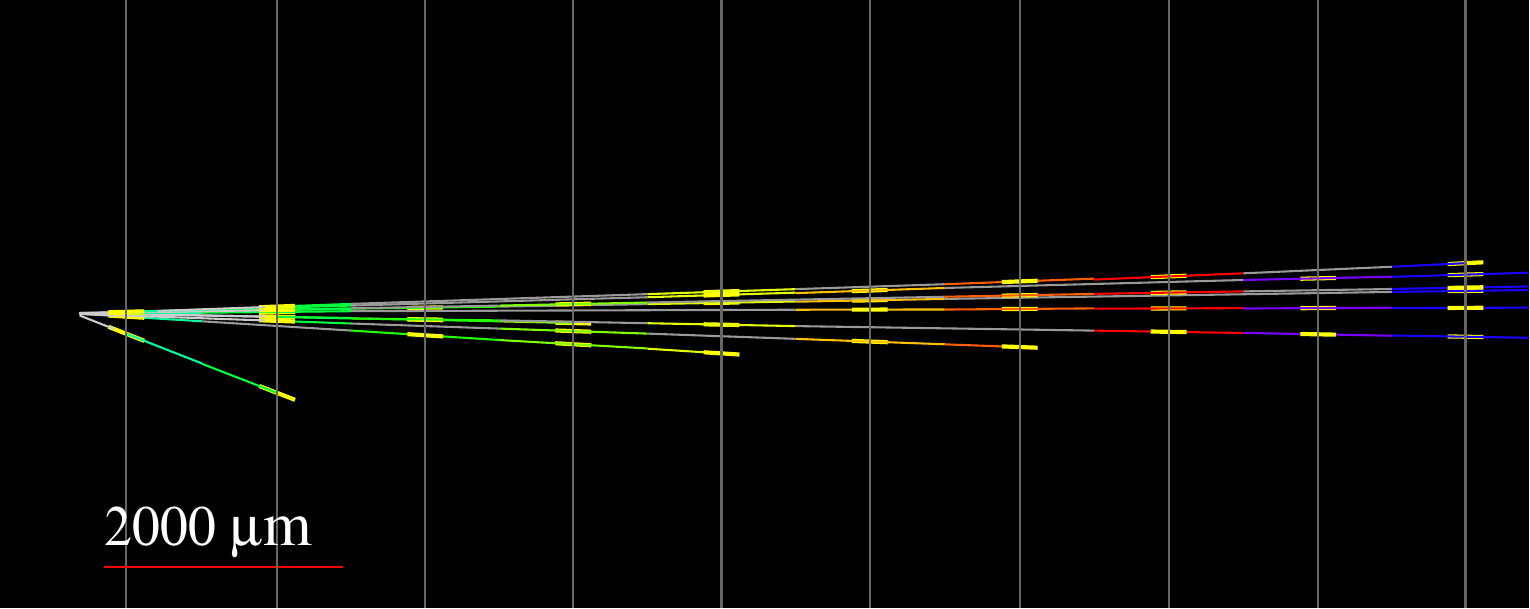}
   \includegraphics[height=0.19\textwidth]{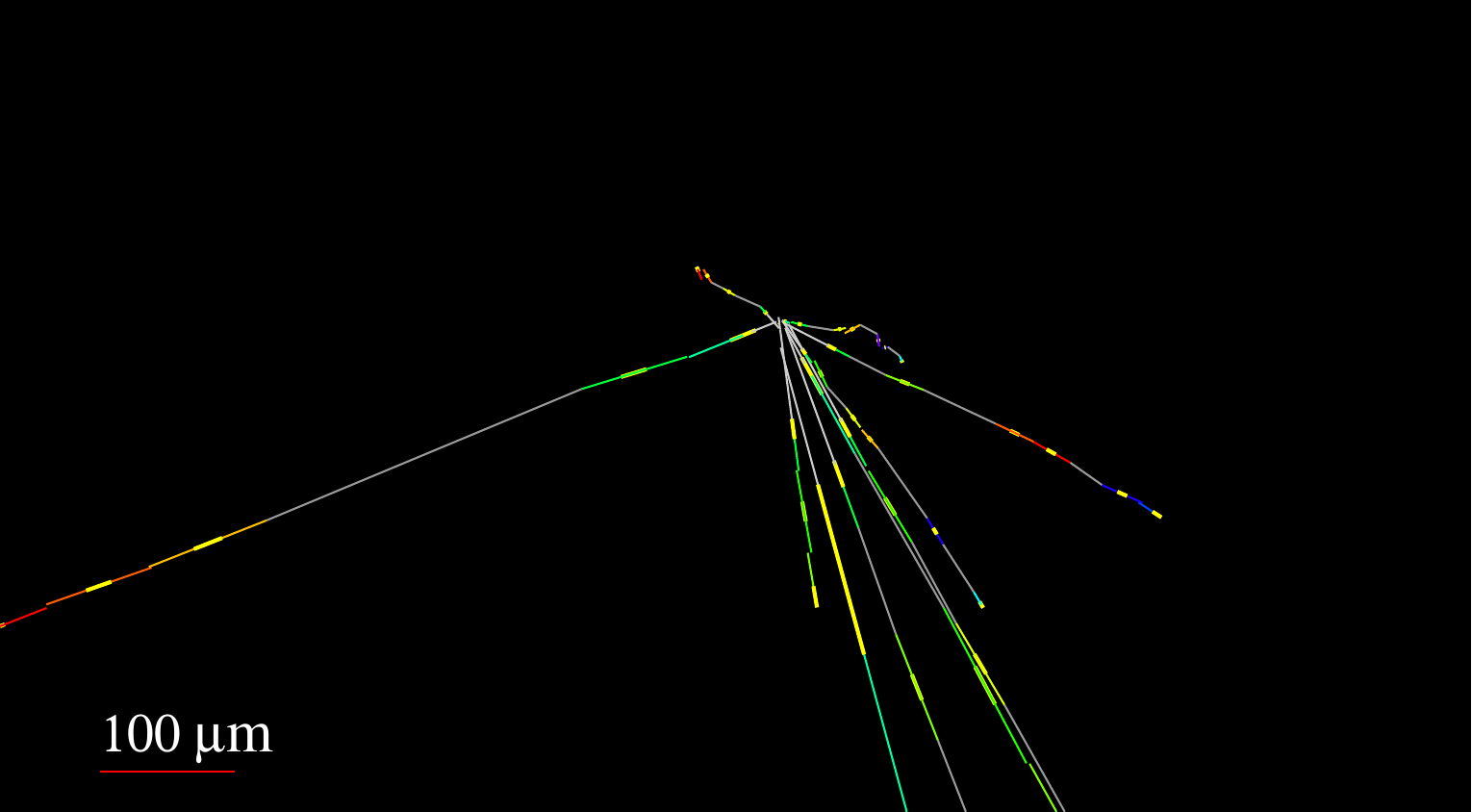}
  \caption{FASER$\nu$ event displays of a neutral vertex in the $y-z$ projection longitudinal to the beam direction (left) and in the view transverse to the beam direction (right).~\cite{FASER:2021mtu}} \label{fg:fasernu} 
\end{figure}

For the HL-LHC, FASER2,~\cite{FASER:2019aik} a larger successor with roughly $R \sim 1~{\rm m}$ and $L \sim 5~{\rm m}$ is planned to be constructed during LS3 and installed in the LHC Forward Physics Facility (FPF) (cf.\ \fref{fg:formosa}). Possible options for FPF locations are an expanded UJ12 cavern or a new cavern $\sim600~{\rm m}$ downstream from IP1. Other experiments proposed to be housed in FPF include FASER$\nu$2,~\cite{fasernu2}, a detector with roughly ten times the mass of FASER$\nu$, as well as the Forward Liquid Argon Experiment (FLArE),\cite{Batell:2021blf} composed of a 10- or 100-tonne-scale liquid argon time projection chamber (LArTPC). 

\subsection{SND@LHC -- Scattering and Neutrino Detector at the LHC}\label{sc:snd}

SND@LHC~\cite{SHiP:2020sos} is a recently approved, compact and stand-alone experiment designed to measure neutrinos produced at the LHC and search for FIPs in the unexplored range of $7.2 < \eta < 8.7$, where neutrinos are mostly produced from charm decay. It is a small-scale prototype of the SHiP~\cite{SHiP:2018yqc} experiment Scattering and Neutrino Detector (SND). The proposed detector is hybrid, combining the nuclear emulsion technology and electronic detector; a schematic view is provided in \fref{fg:snd-lhc}. It will be installed in the TI18 tunnel, in a location off-axis with respect to the ATLAS~\cite{ATLAS:2008xda} IP1. 

\begin{figure}[ht]
\centering
\begin{minipage}[b]{0.48\linewidth}
  \includegraphics[width=\textwidth]{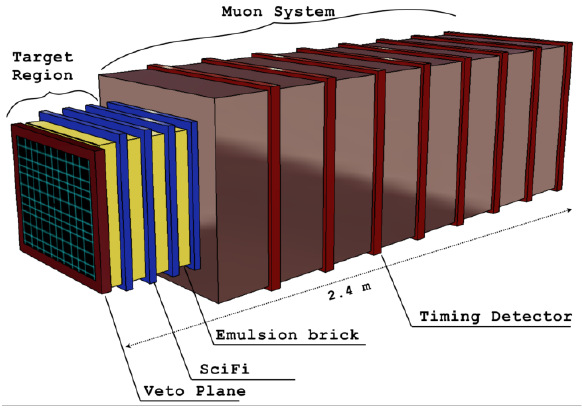}
  \caption{Annotated schematic view of the SND@LHC detector components.~\cite{SHiP:2020sos}} \label{fg:snd-lhc}
\end{minipage}\hspace{0.03\linewidth}
\begin{minipage}[b]{0.48\linewidth}
  \includegraphics[width=\textwidth]{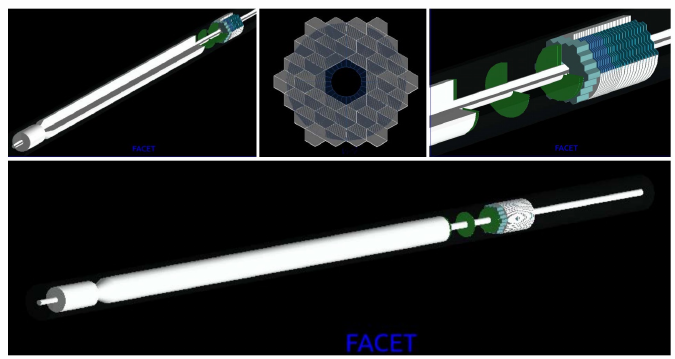}
  \caption{Diagrams of the detector components of the FACET spectrometer.} \label{fg:facet}
\end{minipage}   
\end{figure}

\subsection{FACET -- Forward-Aperture CMS ExTension}\label{sc:facet}

FACET~\cite{facet} is a multi-particle spectrometer to be located at $z \sim +100~{\rm m}$ from the CMS~\cite{CMS:2008xjf} IP5. The detector will have a radius of $\sim 50~{\rm cm}$ and coverage $6 < \eta < 8$, therefore it will be much closer to the IP and feature much larger decay volume than FASER. It is shielded by about $30-50~{\rm m}$ of steel in front of it, which corresponds to $190-300$ interaction lengths, $\lambda_\text{int}$. FACET will be fully integrated in CMS and can be used either a forward part of CMS or a standalone detector. Some {\sc Geant4} modelling drawings of the FACET detector components are shown in \fref{fg:facet}. If approved, FACET will operate during HL-LHC.

\subsection{CODEX-b -- COmpact Detector for EXotics at LHCb}\label{sc:codexb}

CODEX-b~\cite{Gligorov:2017nwh,Aielli:2019ivi} is proposed as a cubic detector with a nominal fiducial volume of $\rm 10~m\times10~m\times10~m$ to be situated in a transverse location $\sim25~{\rm m}$ from the LHCb~\cite{Alves:2008zz} IP8, as depicted in \fref{fg:codex-b}, corresponding to the pseudorapidity range $0.2<\eta <0.6$. It will be composed of six Resistive Plate Chamber (RPC) layers --- being fast, precise and cheap for a large-area tracker --- at $4~{\rm cm}$ intervals on each box face with $1~{\rm cm}$ granularity. An additional passive Pb shielding of $25\lambda_\text{int}$ with embedded active scintillator veto will reduce IP and secondary backgrounds. It is reminded that IP8 runs at a factor of $\sim10$ less luminosity than IP1, IP2 and IP5.

\begin{figure}[ht]
\centering
  \includegraphics[width=0.98\textwidth]{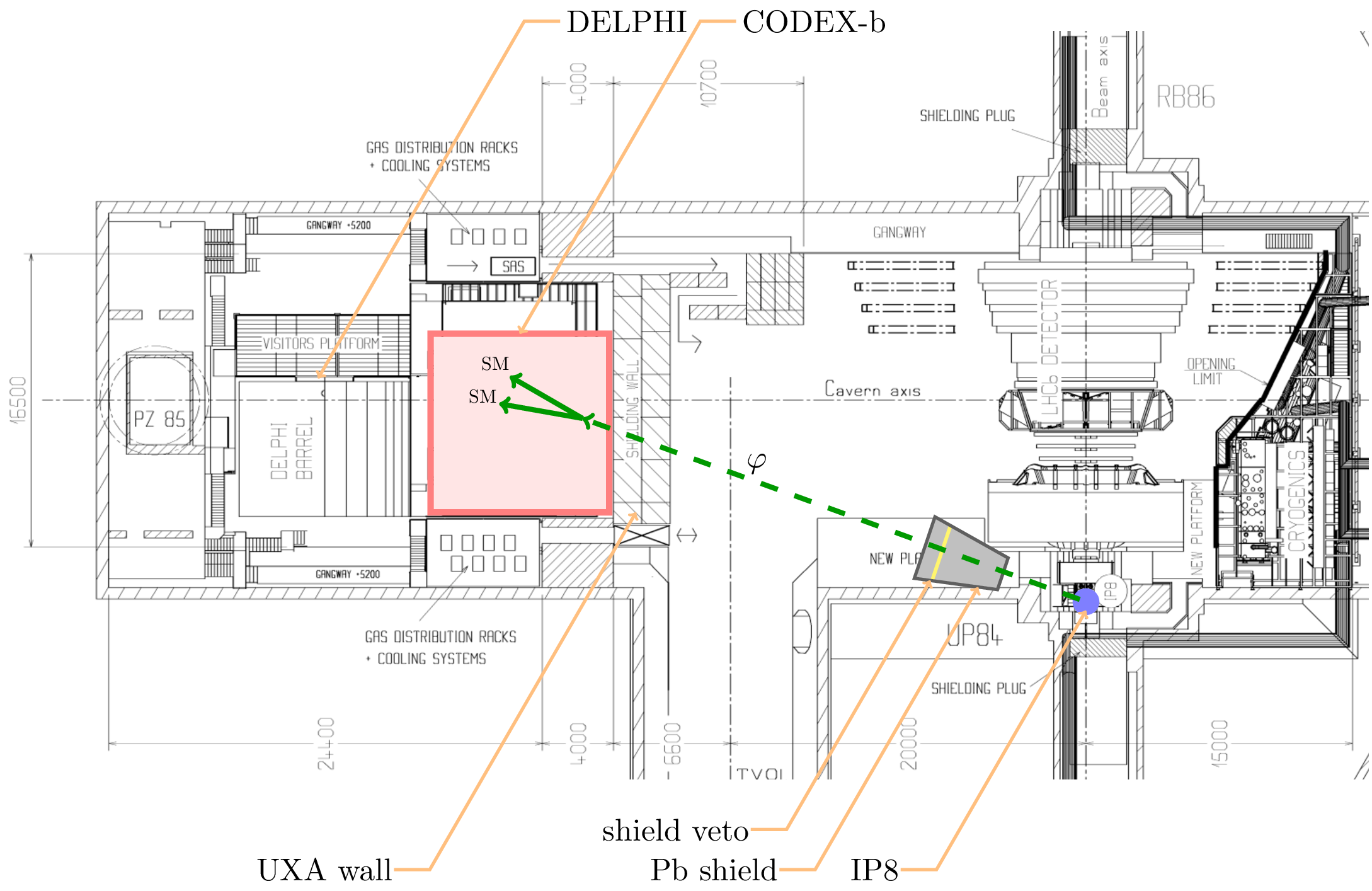}
  \caption{Layout of the LHCb cavern UX85 at IP8, overlaid with the CODEX-b volume.~\cite{Aielli:2019ivi}} \label{fg:codex-b} 
\end{figure}

CODEX-b will be interfaced with LHCb data acquisition, therefore there will have a unique trigger. Further additions to CODEX-b include calorimetry, which would significantly enhance the physics reach, e.g.\ photon signatures. Moreover, absorber or pre-shower layers can also perform particle identification, e.g.\ $e/\gamma$ separation.

The CODEX-$\beta$ demonstrator, with a volume of $2\times 2\times 2~{\rm m}^3$,  is proposed to be operated during Run~3 to provide proof of concept. Its primary goal is to be integrate with LHCb online, reconstruct $K_L^0$ and measure background rates. It will provide competitive sensitivity to $b \to s \chi (\to\text{hadrons})$, as well.

\subsection{MATHUSLA -- MAsive Timing Hodoscope for Ultra Stable neutraL pArticles}\label{sc:mathusla}

MATHUSLA~\cite{MATHUSLA:2019qpy,MATHUSLA:2020uve} is a proposed surface detector of a large footprint (area $100\times 100~{\rm m}^2$) and large decay volume  (height $25~{\rm m}$) to be located above CMS,~\cite{CMS:2008xjf} as depicted in \fref{fg:mathusla}.  The $\sim 90~{\rm m}$ of rock between the IP5 and the detector decay volume provides enough shielding for MATHUSLA to work in a clean environment. The air-filled decay volume will be occupied by several detector layers for tracking in a modular way. RPCs and plastic scintillators are proven technologies that meet the specifications. Being a background-free experiment increases the sensitivity to LLPs up to decay lengths of $10^7~{\rm m}$ and extends the sensitivity of the main detectors by orders of magnitude. 

\vspace*{-0.5cm}

\begin{figure}[ht]
\centering
\begin{minipage}[b]{0.65\linewidth}
\centering
  \includegraphics[width=\textwidth]{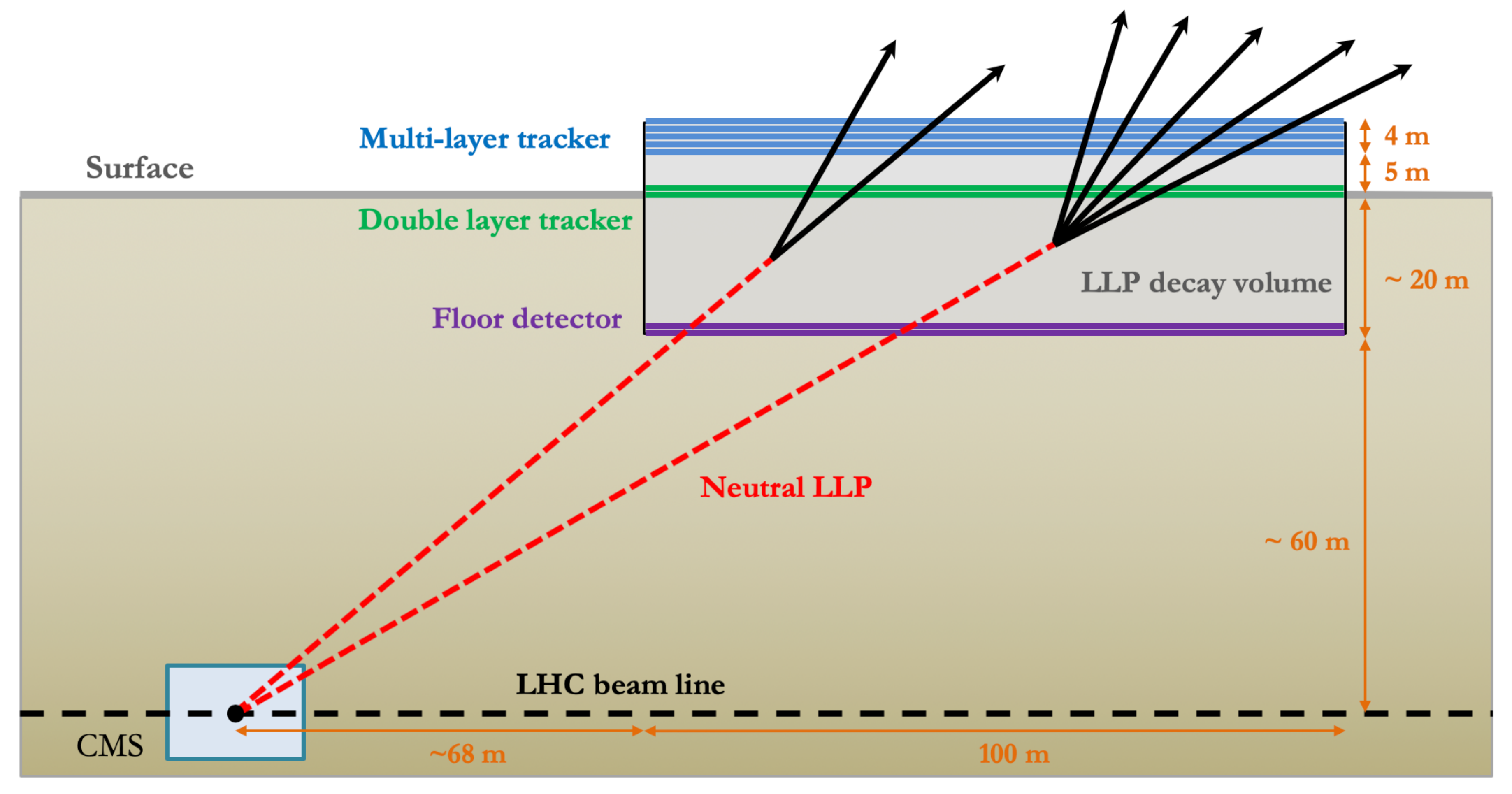}
  \caption{The MATHUSLA detector layout positioned relative to CMS IP5.~\cite{MATHUSLA:2020uve}} \label{fg:mathusla} 
\end{minipage}\hspace{0.02\linewidth}
\begin{minipage}[b]{0.32\linewidth}
\centering
  \includegraphics[width=\textwidth]{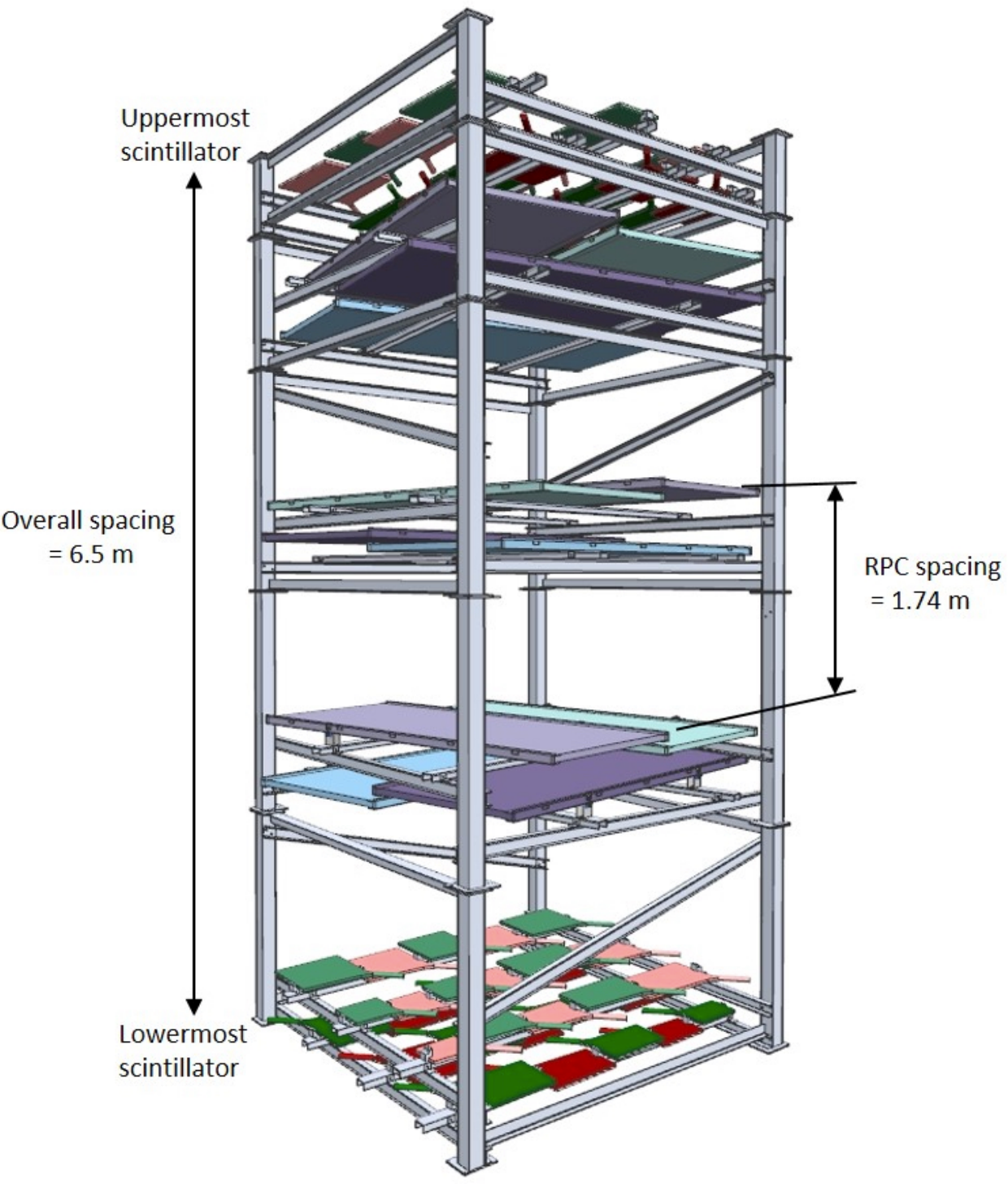}\vspace*{-0.6cm}
  \caption{The MATHUSLA test stand.~\cite{Alidra:2020thg}} \label{fg:mathusla-teststand}
\end{minipage}   
\end{figure}
 
A test stand following the concept of the full detector was installed in the surface area above the ATLAS~\cite{ATLAS:2008xda} IP, taking data with different beam conditions during 2018.~\cite{Alidra:2020thg} The $2.5\times 2.5\times 6.5~{\rm m}^3$ unit detector is composed of one external layer of scintillators in the upper part and one in the lower part with six layers of RPCs between them, as shown in \fref{fg:mathusla-teststand}. The obtained results confirmed the background assumptions in the MATHUSLA proposal.

Moreover, MATHUSLA could act as a cosmic-ray (CR) telescope performing very precise measurements up to the PeV scale.~\cite{Curtin:2018mvb,Alpigiani:2020iam} By integrating a device with the possibility to measure arrival times and particle densities of extensive air showers, such as an RPC, MATHUSLA can be employed as a CR detector and monitor a big portion of the sky above $(\theta < 80^{\circ})$, without limitation to inclined events.

\subsection{ANUBIS -- AN Underground Belayed In-Shaft}\label{sc:anubis}

ANUBIS~\cite{Bauer:2019vqk} is an off-axis detector designed for neutral LLPs with $c\tau \gtrsim 5~{\rm m}$, proposed to occupy the PX14 installation shaft of the ATLAS~\cite{ATLAS:2008xda} experiment (cf.\ \fref{fg:anubis}(left)), which is not used during regular LHC operation. It will comprise four evenly spaced tracking stations (TS) with a cross-sectional area of $230~{\rm m}^2$ each, shown in \fref{fg:anubis}(right). The tracking stations will use the same RPC technology as the new ATLAS layers and ATLAS can be used as an active veto of SM activity. The projective decay volume optimises the acceptance for different lifetimes. Two smaller TS prototypes are planned to be installed for Run~3 as a demonstrator. 

\begin{figure}[ht]
\centering
\begin{minipage}[b]{0.44\linewidth}\centering
  \includegraphics[width=0.95\textwidth]{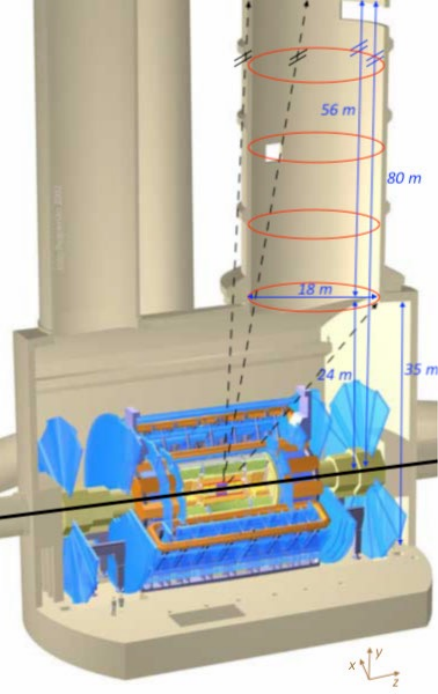}
\end{minipage}\hspace{0.03\linewidth}
\begin{minipage}[b]{0.52\linewidth}
\centering
  \includegraphics[width=0.9\textwidth]{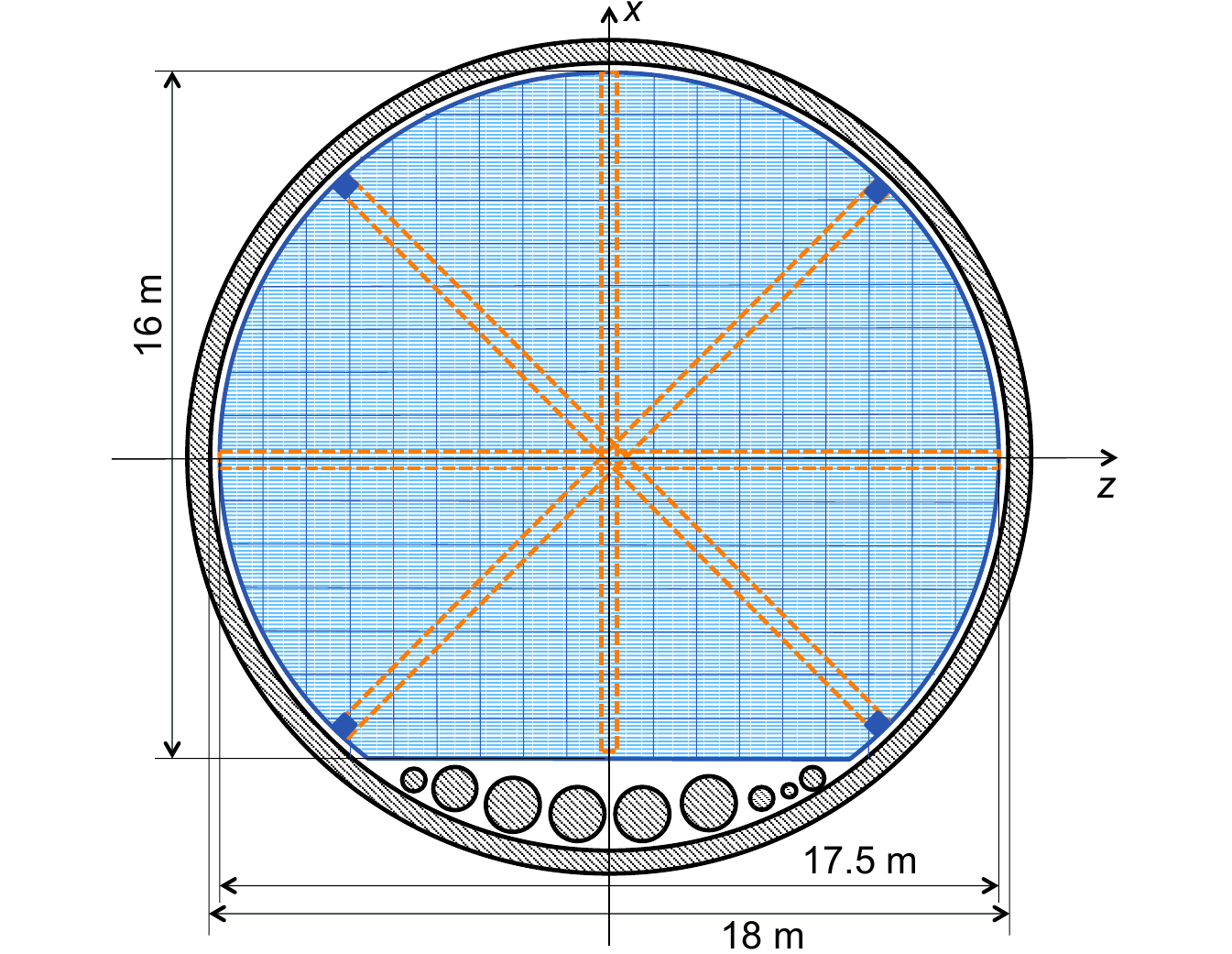}
  \caption{Left: Location of the ANUBIS detector in the PX14 installation shaft of the ATLAS experiment. Above: One of the four ANUBIS tracking stations in the $(x, z)$ plane. The shaft walls and the ATLAS cavern pipework are shown in gray, the TS in blue and the TS support structure in orange.~\cite{Bauer:2019vqk}} \label{fg:anubis}
\end{minipage}   
\end{figure}

\subsection{AL3X -- A Laboratory for Long-Lived eXotics}\label{sc:al3x}

In the event that ALICE~\cite{ALICE:2008ngc} finishes its physics program before the end of HL-LHC, it has been proposed~\cite{Gligorov:2018vkc} to reuse the L3 magnet at IP2 and the ALICE Time Projection Chamber (TPC) for LLP searches. This proposal requires to move the IP by $11.25~{\rm m}$ outside the magnet, as shown in \fref{fg:al3x}, to allow LLPs to travel before decaying. An additional thick shield and an active veto ($D_1$ and $D_3$) will reduce the background. The detector geometry corresponds to a pseudorapidity acceptance $0.9 < \eta < 3.7$. The AL3X configuration is in essence a tracking detector behind a heavy shield, which can be thought of as analogous to a calorimeter that is solely absorber. This permits AL3X to search for much rarer signals in a very low background environment compared to ATLAS and CMS, and in this sense AL3X would be complementary to the existing (and proposed upgraded) multi-purpose detectors.

\begin{figure}[ht]
\centering
  \includegraphics[width=0.8\textwidth]{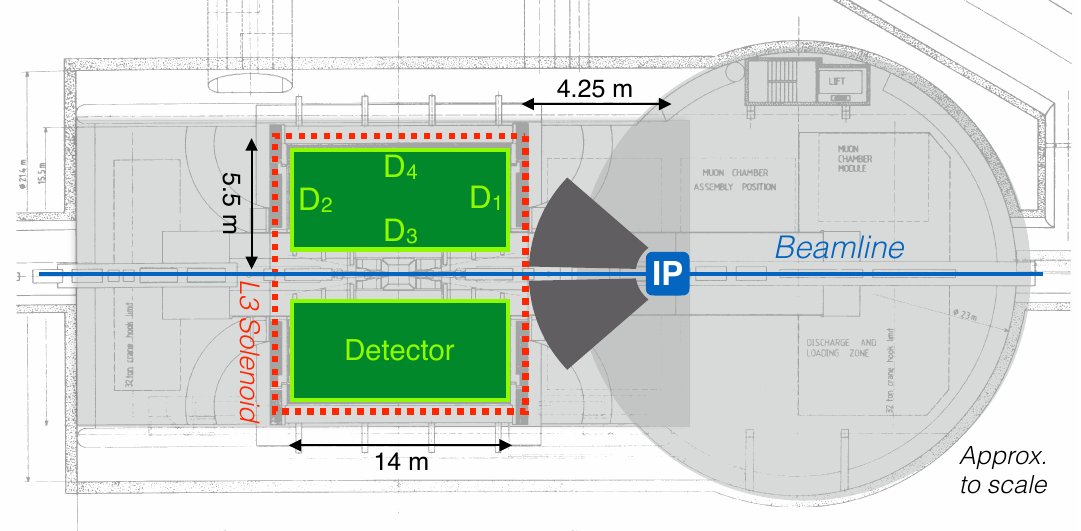}
  \caption{Schematic view of the AL3X detector layout: the cylindrical TPC (dark green) and the veto and trigger layers $D_i$ (light green). The current L3 magnet is shown (dashed red) for reference.~\cite{Gligorov:2018vkc}} \label{fg:al3x} 
\end{figure}

\subsection{milliQan}\label{sc:milliqan}

Just like MAPP-mQP, the milliQan detector,~\cite{Haas:2014dda,Ball:2016zrp} to be installed near the CMS~\cite{CMS:2008xjf} IP5, will search for stable \emph{millicharged} particles, unlike the majority of the aforementioned detectors which are sensitive to visible SM particles originating from \emph{neutral} DM decays. The milliQan is located in an underground tunnel at a distance of $33~{\rm m}$ from the CMS IP, with $17~{\rm m}$ of rock between the IP and the detector that provides shielding from most particles produced in LHC collisions. In order to be sensitive to particles with charges as low as $0.001e$, a large active area of scintillator is required. For Run~3, two detector designs are planned for deployment. The bar detector is made of ${\rm 0.2~m \times 0.2~m  \times 3~m}$ plastic scintillator bars surrounded by an active $\mu$ veto shield. The slab detector will increase the reach for heavier mCPs through ${\rm 40~cm \times 60~cm \times 5~cm}$ scintillator slabs.~\cite{milliQan:2021lne}

A small fraction of $\sim1\%$ of the full detector, the \emph{milliQan demonstrator,} shown in \fref{fg:milliqan} was installed and operated to measure backgrounds and provide proof of principle. It allowed the first search for millicharged particles at a collider.~\cite{Ball:2020dnx} A data sample of 37.5~\ifb $pp$ collisions at $\sqrt{s}=13~\tev$ has been analysed and no excess over the background prediction observed. The results interpretation in terms dark photons is discussed later in \sref{sc:mcp}.

\begin{figure}[ht]
\centering
\begin{minipage}[b]{0.48\linewidth}
  \includegraphics[width=0.9\textwidth]{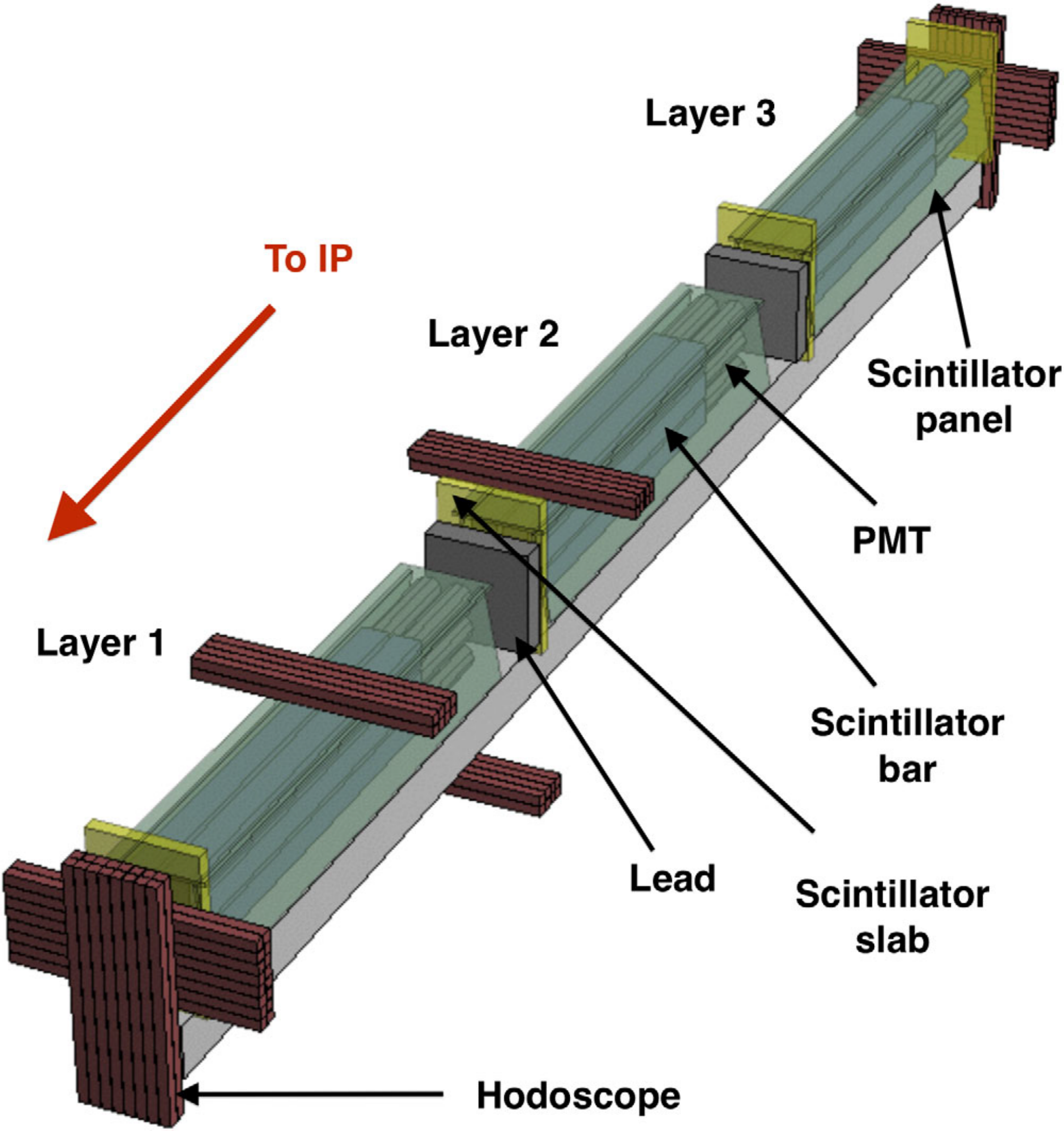}
  \caption{Diagram of the milliQan demonstrator components.~\cite{Ball:2020dnx}} \label{fg:milliqan}
\end{minipage}\hspace{0.03\linewidth}
\begin{minipage}[b]{0.48\linewidth}
  \includegraphics[width=\textwidth]{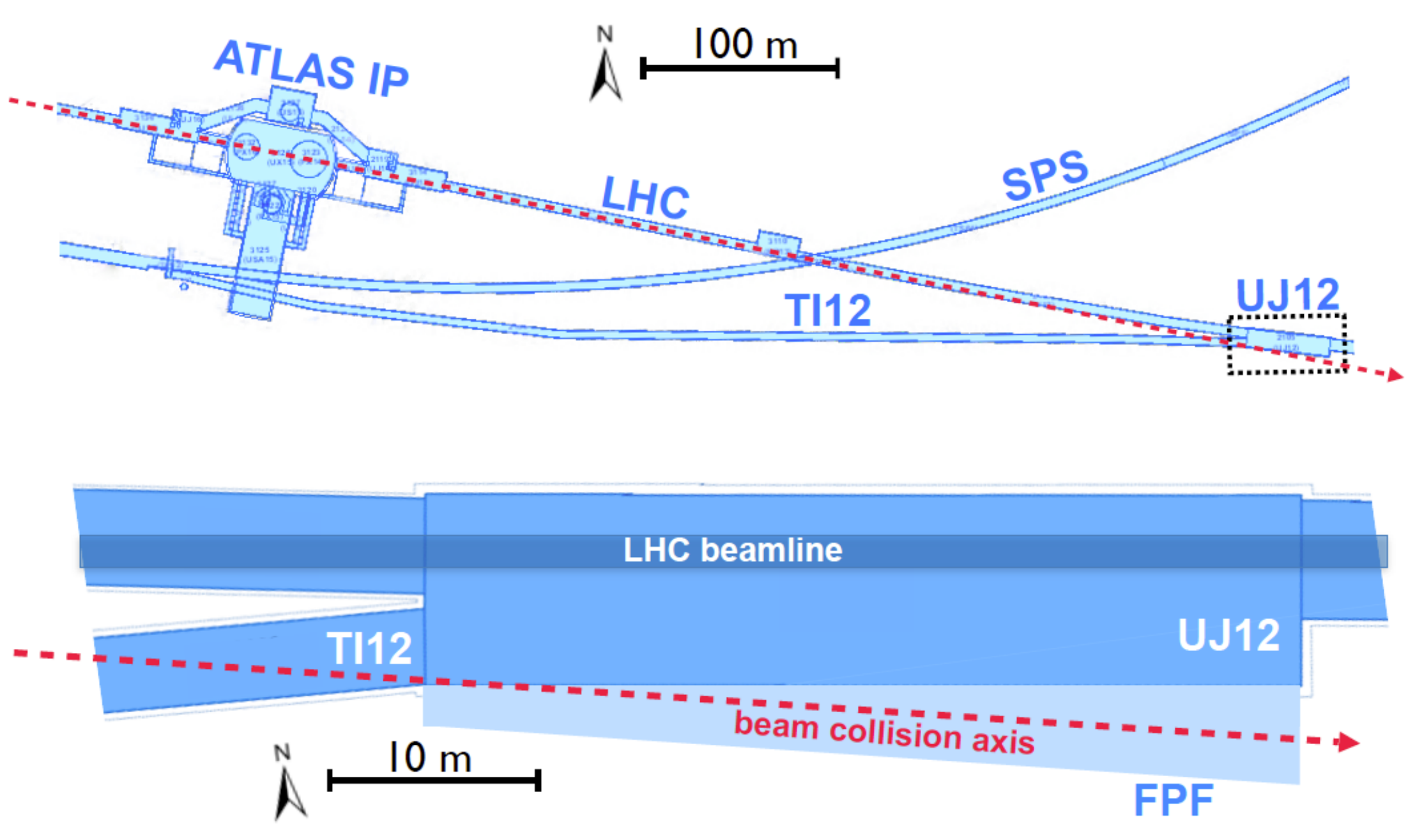}
  \caption{Location of FORMOSA in the cavern UJ12 or side tunnel TI12 (blue) close to the beam collision axis (red). The FPF extension is shown as a light-blue area.~\cite{Foroughi-Abari:2020qar}} \label{fg:formosa}
\end{minipage}
\end{figure}

\subsection{FORMOSA -- FORward MicrOcharge SeArch}\label{sc:formosa}

FORMOSA~\cite{Foroughi-Abari:2020qar} is also designed to discover millicharged particles, yet in the far-forward direction, close to the beam collision axis, where it can benefit from an enhanced mCP production cross section compared to the transverse direction. It is a scintillator-based detector to be hosted in the FPF, shown in Fig.~\ref{fg:formosa}. FORMOSA is proposed to start in Run~3 by moving the milliQan demonstrator to UJ12, as a phase called FORMOSA-I, with the full milliQan-type detector, FORMOSA-II, to be deployed at a later stage. The expected physics potential is discussed in \sref{sc:mcp}.

\section{The Higgs Portal}\label{sc:higgs}

The so-called \emph{scalar portal} involves a new dark scalar $S$, which mixes slightly with the SM-like Higgs, and provides a simple target for new physics searches.~\cite{Patt:2006fw} The dark Higgs mixing portal admits exotic inclusive $B \to X \phi$ decays,\footnote{$D$ mesons and kaons have much smaller branching ratios into a Higgs-mixed scalar and are neglected.} where $\phi$ is a light CP-even scalar that mixes with the SM Higgs, with a mixing angle of $\theta \ll 1$.  The particle lifetime depends on the degree of mixing. One possible, simple Lagrangian which includes this new dark Higgs mixing is given by~\cite{FASER:2018eoc}
\begin{equation}
        \mathcal{L} = \mathcal{L}_\text{SM} + \mathcal{L}_\text{DS}+ \mu_S^2 S^2 - \frac{\lambda_S}{4} S^{4} - \epsilon S^2 H^{\dagger}H,
\end{equation}
where $S$ is a real scalar field, $H$ is a SM-like Higgs field, $\epsilon$ is the portal coupling, and $\lambda_S$ is a free parameter.  The quartic term contains the mixing between the SM Higgs and the new scalar, with the resulting physical fields:  the SM Higgs $h$ and the dark Higgs $\phi$.  Both fields acquire a non-zero VEV and the coupling between these two particles induces new Yukawa-like couplings between the dark Higgs and the SM fermions. In addition, there can appear a non-negligible trilinear interaction term between $\phi$ and $h$ with the corresponding coupling denoted by $\lambda$, i.e.\ $\propto\lambda h\phi\phi$.  Thus, the signal sought after in LLP experiments is two charged lepton tracks originating from dark Higgs decays $\phi \to \ell^{+}\ell^{-}$, in their fiducial volume.  

Currently the best experimental limits on dark Higgs production at colliders come from CHARM~\cite{CHARM:1985anb} and LHCb,~\cite{LHCb:2015nkv,LHCb:2016awg} shown in \fref{fg:mapp-higgs} in the $\sin^2\theta$-versus-$m_S$ plane. If no trilinear coupling is assumed and the background is considered negligible, we derive the exclusion curves for MATHUSLA, SHiP, CODEX-b, MAPP-1 and MAPP-2 that are shown in the same figure. MAPP-1 during Run~3 (30~\ifb) is expected to provide considerable coverage, when the other experiments will be under construction; from the CODEX-b side, the CODEX-$\beta$ demonstrator will be in operation. MAPP-2 extents significantly MAPP-1 sensitivity making it competitive with SHiP. As expected, MATHUSLA has better sensitivity in lower mixing angles, thus longer lifetimes, due to its large distance from the IP.

\begin{figure}[ht]
\centering
\begin{minipage}[b]{0.55\linewidth}
  \includegraphics[width=\textwidth]{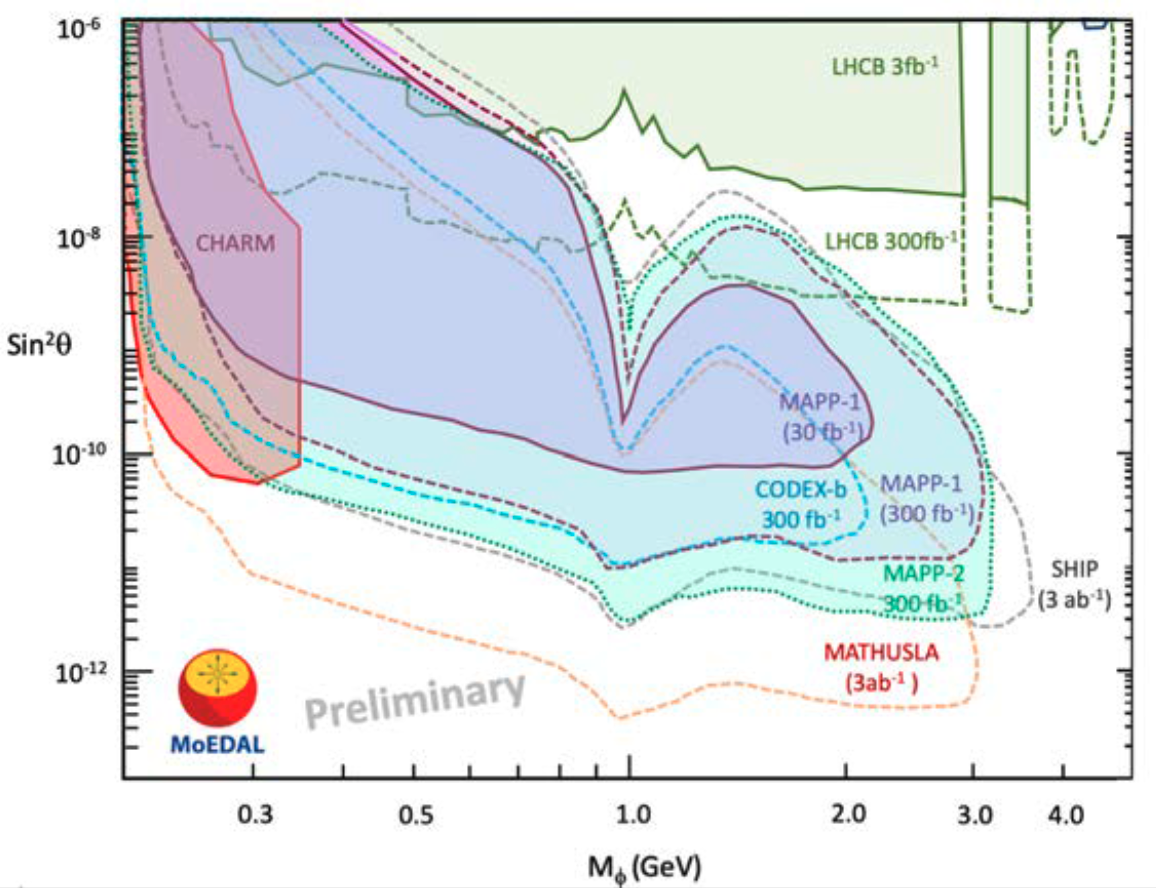}
  \end{minipage}\hspace{0.03\linewidth}
 \begin{minipage}[b]{0.41\linewidth}
\caption{95\% CL exclusion bounds Projected sensitivity of MoEDAL-MAPP and other experiments for dark Higgs bosons produced in rare $B$~decays at $\sqrt{s}=14~\tev$. Adopted from Ref.~\citenum{Gligorov:2018vkc}. \\ } \label{fg:mapp-higgs}
\end{minipage}
\end{figure}

Assuming now a non-vanishing trilinear coupling $\lambda$, the expected sensitivity from the experiments presented in \sref{sc:experiments} is shown in \fref{fg:pbc-higgs}. This time, a larger fraction of the parameters space is covered due to the contributions arising from a Higgs in the decay cascade, either virtual, e.g.\ the $B \to K S S$ mode, or real, e.g.\ the $h \to S S$. The branching fraction $BR(h \to S S)\sim 10^{-2}$ is considered to remain compatible with the LHC searches for the Higgs to invisible channels. The larger impact is provided by the bigger experiments, MATHUSLA,~\cite{Evans:2017lvd} SHiP,~\cite{Alekhin:2015byh} FASER2~\cite{FASER:2018eoc} and CODEX-b,~\cite{Beacham:2019nyx} which can explore the region well above the \gev mass scale in a fully uncharted range of couplings. 

\begin{figure}[ht]
\centering
  \includegraphics[width=0.87\textwidth]{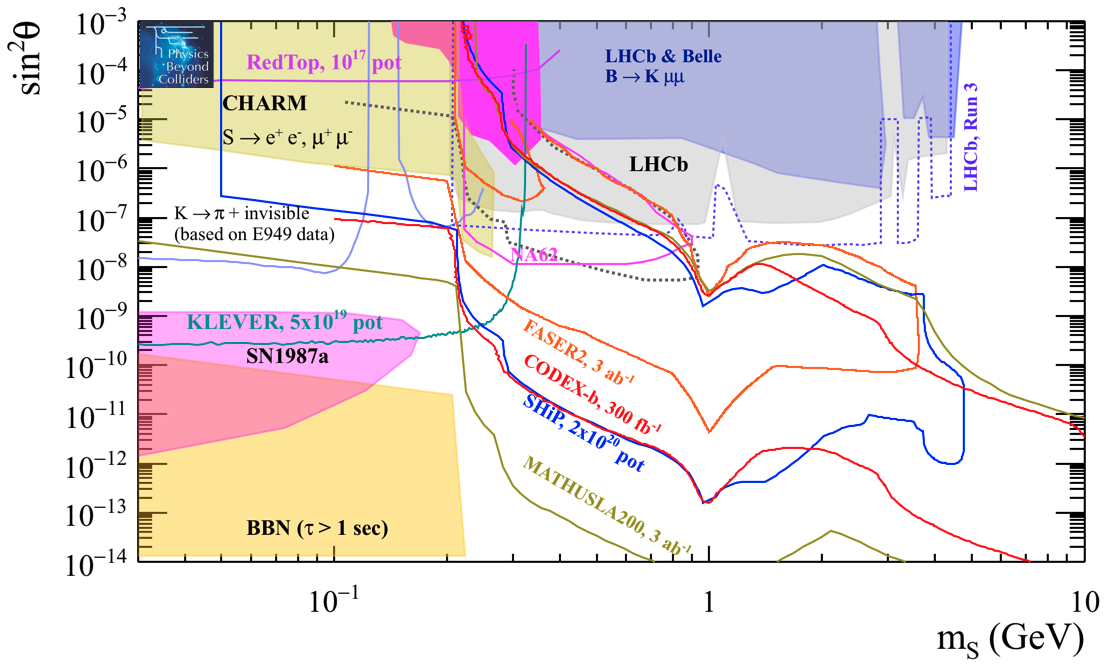}
  \caption{Prospects on 10-15 year timescale for dark-scalar mixing with Higgs in the  $(m_S,\sin^2\theta)$ plane with trilinear coupling $\lambda \neq 0$ and assuming $BR(h \to SS) = 10^{-2}$. The NA62$^{++}$ and KLEVER curves correspond to $\lambda=0$, so they are conservative.~\cite{Beacham:2019nyx}} \label{fg:pbc-higgs}
\end{figure}

Dark scalar portals may have cosmological implications, contributing to the observed DM abundance. Dark Higgs bosons may mediate interactions with hidden dark matter that has the correct thermal relic density or resolves small-scale-structure discrepancies. Indeed, there are models with a dark Higgs inflaton strongly favoured by cosmological Planck~\cite{Planck:2018jri} and BICEP/Keck Array~\cite{BICEP2:2018kqh} data that constrain the energy scale of
inflation. One of these is expected to leave imprints on LHC experiments, such as FASER and MAPP-1.~\cite{Popa:2021fgy}

\section{Dark Photons: Neutral Metastable States}\label{sc:dp}

A large class of BSM models includes interactions with light new \emph{vector} particles. Such particles can result from additional gauge symmetries of BSM physics. New vector states can mediate interactions between the SM fields and extra dark-sector fields that may eventually play the role of the DM states. 

The most minimal vector portal interaction can be written as 
\begin{equation}
\label{vector}
{\cal L}_\text{vector} = {\cal L}_{\rm SM} + {\cal L}_{\rm DS} -\frac{\epsilon}{2\cos\theta_W} F'_{\mu\nu} B^{\mu\nu},
\end{equation}
where ${\cal L}_{\rm SM}$ is the SM Lagrangian, $B_{\mu\nu}$ and $F'_{\mu\nu}$ are the field stengths of hypercharge and new $U'(1)$ gauge groups, $\epsilon$ is the so-called \emph{kinetic mixing} parameter,~\cite{Holdom:1985ag} and  ${\cal L}_{\rm DS}$ stands for the dark sector Lagrangian that may include new matter fields $\chi$ charged under $U'(1)$,
\begin{equation}
{\cal L}_{\rm DS} = -\frac{1}{4} F'_{\mu\nu} F'^{\mu\nu}  + \frac{1}{2}m_{A'}^2 A'_\mu A'^\mu +   |(\partial_\mu + ig_DA'_\mu )\chi|^2 +...
\end{equation}
If $\chi$ is (meta)stable, it may constitute a fraction or entirety of dark matter. At low energy this theory contains a new massive vector particle, a \emph{dark photon} state, coupled to the electromagnetic current with $\epsilon$-proportional strength, $\epsilon A'_{\mu}   J^\mu_\text{EM}$. 

In the minimal dark photon model, DM is assumed to be either heavy or contained in a different sector. The dark photon, $\gamma_{\rm d}$, decays into SM states (visible decays). The physics potential of the proposed LLP experiments as a function of the  dark photon mass $m_{A'}$ and the coupling of dark photon with the SM photon $\epsilon$ is shown in \fref{fg:dark-photon}, compared with existing bounds from several beam-dump data, e.g.\ Refs.~\citenum{Blumlein:2011mv,Blumlein:2013cua}. The sensitivity for dark photons decaying into visible final states is expected to be dominated by SHiP,~\cite{Alekhin:2015byh} while FASER2,~\cite{FASER:2018eoc} LDMX~\cite{Berlin:2018bsc} and AWAKE~\cite{Caldwell:2018atq} will be directly competing with SeaQuest, LHCb, Heavy Photon Search (HPS)~\cite{HPS:2018xkw} experiment, and others. MATHUSLA in this scenario is however not competitive, mostly due to the fact that the dark photon is produced in the forward direction. 

\begin{figure}[htb]
   \centering
   \includegraphics[width=0.85\linewidth]{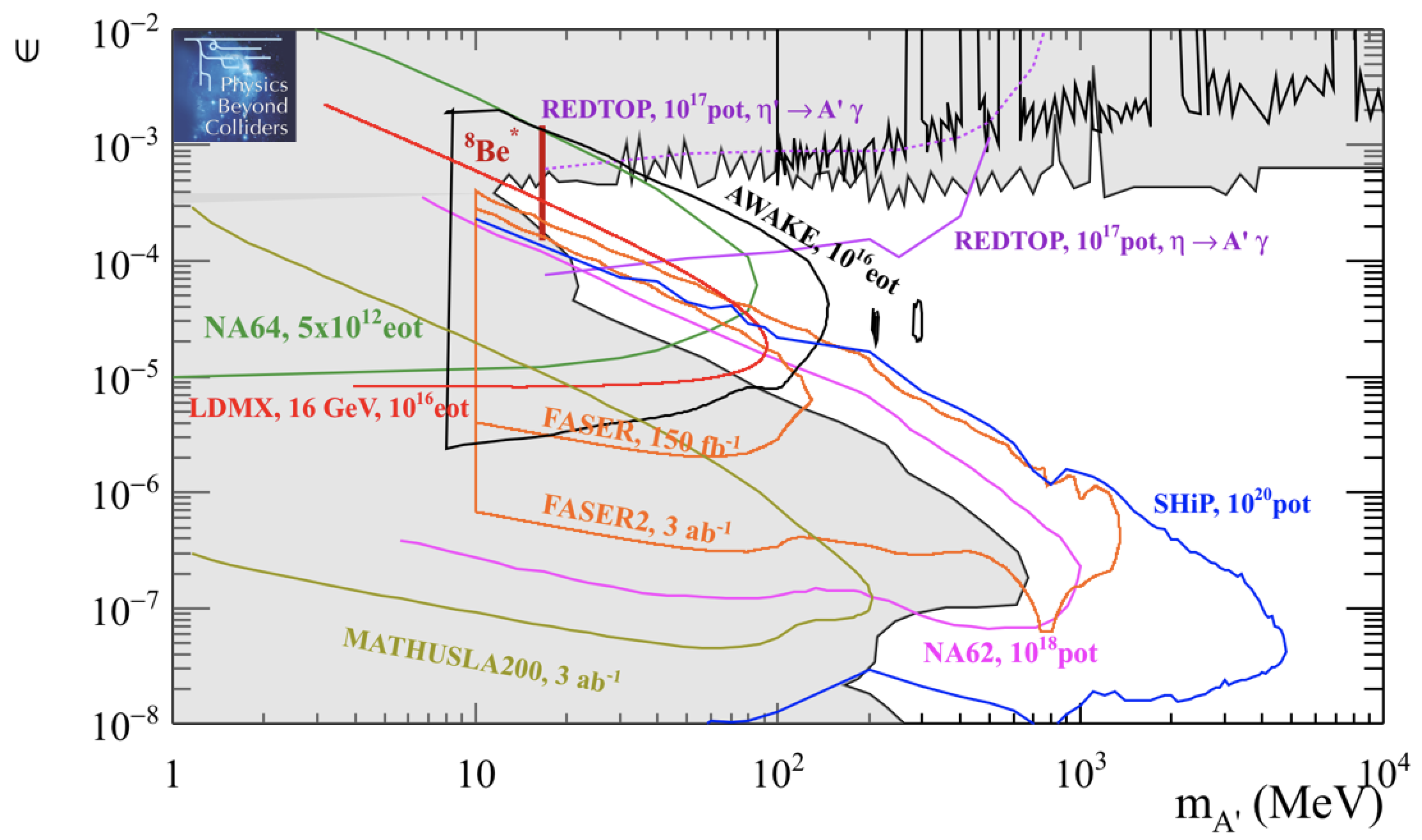}
   \caption{Future upper limits at 90\% CL for a minimal dark-photon model with visible decays in the plane mixing strength $\epsilon$ versus mass $m_{A'}$ for various projects on a $\sim$10--15 year timescale~\cite{Beacham:2019nyx}. }
   \label{fg:dark-photon}
\end{figure}

In a different scenario, light (sub-\gev) DM, $\chi$, may be coupled to a dark photon, constituting a minimally coupled weakly interacting massive particle (WIMP) model. The preferred values of the coupling $g_D$ are such that $A'\to \chi\chi$, with $\chi$ potentially scattering further on electrons and nuclei, while $m_{A'}/m_{\chi}=3$ is assumed for the masses. Recent studies on DM scattering off electrons~\cite{Batell:2021blf} and off nuclei~\cite{Batell:2021aja} in an emulsion detector (FASER$\nu$2) and a LArTPC  (FLArE) in the FPF showed very interesting sensitivity, probing the thermal-relic region. Neutrino background may be separated from DM signal using energy and angle selection criteria. Such forward detectors open the possibility for performing a direct-detection-type DM search at the LHC. 

In inelastic DM, which constitutes a viable and compelling paradigm for light thermal DM, DM couples to the SM only by interacting with a nearly degenerate dark-photon mediator heavier than $\sim10~\gev$. For relative mass-splittings larger than ${\cal O}(10^{-6})$, DM-nucleon/electron scattering at direct detection experiments is kinematically suppressed. However, at the LHC, where the DM and excited state can be directly produced, for mass-splittings above a few \mev, the excited state can decay back to DM and a pair of SM fermions, often on collider timescales. giving rise to visible displaced vertices. ATLAS, CMS, LHCb, CODEX-b, FASER, and MATHUSLA can detect such DM signals in the cosmologically motivated mass range of $\sim1-100~\gev$.~\cite{Berlin:2018jbm}

\section{Dark Photons: Stable Millicharged Particles}\label{sc:mcp}

Millicharged particles have been discussed in connection with the mechanism of electric charge quantisation and possible non-conservation of electric charge.\cite{Ignatiev:1978xj} There are three experiments planned to run at the LHC that are sensitive to the detection of the low ionisation coming from an mCP: milliQan (\sref{sc:milliqan}), the MAPP-mQP sub-detector (\sref{sc:mapp}) and, the recently proposed, FORMOSA (\sref{sc:formosa}). 

A well-motivated mechanism that predicts mCPs is the introduction of a new massless $U'(1)$ gauge field, the dark photon, $A'_{\mu \nu}$, which is coupled to the SM photon field, $B^{\mu \nu}$.  A new massive dark fermion $\psi$ \emph{(dark QED)} of mass $M_\text{mCP}$, is predicted, which is charged under the new $U'(1)$ field $A'$ with charge $e'$. The Lagrangian for the model is given by~\cite{Haas:2014dda}
\begin{equation}
    \mathcal{L} = \mathcal{L}_\text{SM} - \frac{1}{4} A'_{\mu \nu}A'^{\mu \nu} + i\bar{\psi}(\slashed{\partial} +ie'\slashed{A'} + iM_\text{mCP})\psi -\frac{\kappa}{2}A'_{\mu \nu} B^{\mu \nu}.
    \label{eq:mcp1}
\end{equation}
The last term contains the kinetic mixing, which one can eliminate by expressing the new gauge boson as, $A'_{\mu} \to  A'_{\mu} + \kappa B_{\mu}$.  Applying this field redefinition reveals a coupling between the charged matter field $\psi$ to the SM hypercharge. The Lagrangian \eqref{eq:mcp1} then becomes:
\begin{equation}
    \mathcal{L} = \mathcal{L}_\text{SM} - \frac{1}{4} A'_{\mu \nu}A'^{\mu \nu} + i\bar{\psi}(\slashed{\partial} +ie'\slashed{A'} -i\kappa e' \slashed{B}  +iM_\text{mCP})\psi.
        \label{eq:mcp2}
\end{equation}

It is now apparent that the field $\psi$ acts as a field charged under hypercharge with a millicharge $\kappa e'$, which couples to the photon and $Z^0$ boson with a charge $\kappa e'\cos{\theta}_{W}$ and $- \kappa e' \sin{\theta}_{W}$, respectively. Expressing the fractional charge in terms of electric charge thus gives $\epsilon = \kappa e' cos{\theta}_{W}/e$.

\begin{figure}[ht]
\centering
\begin{minipage}[b]{0.48\linewidth}
  \includegraphics[width=0.98\textwidth]{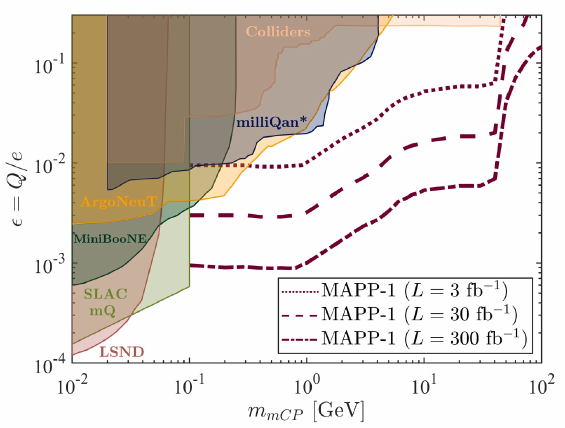}
  \caption{95\% CL exclusion limits for dark fermion mCPs in the mass vs.\ charge plane obtained by the milliQan demonstator~\cite{Ball:2020dnx} compared to previous constraints and to the MAPP-mQP sensitivity  for various integrated luminosity assumptions.~\cite{Staelens:2021}} \label{fg:dark-photon-mcp}
\end{minipage}\hspace{0.03\linewidth}
\begin{minipage}[b]{0.48\linewidth}
  \includegraphics[width=0.97\textwidth]{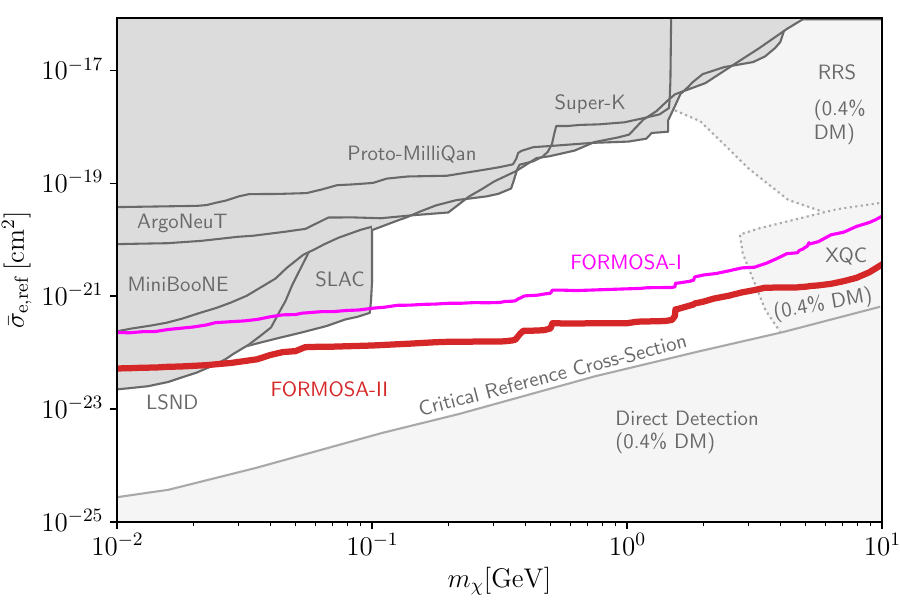}
  \caption{Sensitivity reaches of FORMOSA in the millicharged SIDM window in terms of the reference cross-section $\bar{\sigma}_{\rm e,ref}$. In addition to accelerator constraints, constraints from direct-detection experiments (assuming $0.4\%$ DM abundance for the direct-detection experiments) are drawn.~\cite{Foroughi-Abari:2020qar}} \label{fg:formosa-ref-xsec}
\end{minipage}   
\end{figure}

The milliQan 1\% scale demonstrator discussed in \sref{sc:milliqan} and shown in \fref{fg:milliqan} provided the first constraints on mCPs at LHC, as shown in \fref{fg:dark-photon-mcp}. Pair-production of millicharged particles at 13~\tev is considered through the Drell-Yan process, as well as from $\Upsilon$, $J/\psi$, $\psi$(2S), $\phi$, $\rho$, and $\omega$ decays into mCP pairs, and from Dalitz decays of $\pi^0$, $\eta$, $\eta'$, and $\omega$. This search excluded mCP masses of $20-4700~\mev$ for charges varying between $0.006e$ and $0.3e$ depending on mass.~\cite{Ball:2020dnx} MAPP-mQP, on the other hand, can extend these limits especially towards higher masses, while the reach in terms of low charges depends on the integrated luminosity; charges as low as $10^{-3}e$ will need the HL-LHC,~\cite{Staelens:2021} as evident from \fref{fg:dark-photon-mcp}. Similar sensitivity is expected by the full milliQan detector during the HL-LHC run.\cite{milliQan:2021lne} 

When considering DM abundance, mCPs can account for a fraction of it that cannot be detected by direct-detection experiments due to attenuation effects, when the ambient DM has substantial cross section with SM particles making it to  lose most of its kinetic energy. For some model parameters,~\cite{Emken:2019tni, Plestid:2020kdm} the DM particles cannot be detected by ground-based direct detection experiments after interacting with the atmospheric particles and the crust. These DM particles are generally referred to as strongly interacting DM (SIDM). In general, mCP-hunting experiments constrain the larger cross sections, leaving a gap, that can be filled by FORMOSA.~\cite{Foroughi-Abari:2020qar} This is shown in \fref{fg:formosa-ref-xsec}, where the accessible \emph{reference cross section} versus mCP mass is drawn for accelerator and for direct-detection experiments, together with the FORMOSA expected reach. 

In addition to DM, apparent fractionally charged particles may arise as heavy neutrinos with a large enough electric dipole moment (EDM) to yield ionisation in mCP-sensitive detectors. It has been demonstrated that milliQan and MAPP-mQP have very good sensitivity in a scenario where the heavy neutrino is considered to be a member of a fourth generation lepton doublet with the EDM introduced within a dimension-five operator.~\cite{Sher:2017wya,Frank:2019pgk}

\section{More Portals and Beyond}\label{sc:other}

As highlighted in the introduction in \sref{sc:intro}, besides the hidden sectors directly related to DM, namely the dark Higgs and dark photons, there are other portals, such as those predicting heavy neutral leptons and axion-like particles (ALPs). The quest for these objects is of equal interest to Particle Physics open questions, they are often related to dark matter and are similarly relevant to the above mentioned LLP experiments, as those discussed in Secs.~\ref{sc:higgs},~\ref{sc:dp} and~\ref{sc:mcp}. Besides the portal connection to DM, LLPs can probe specific DM models, such as the case of MATHUSLA and dynamical dark matter.~\cite{Curtin:2018ees}

\subsection{Extended neutrino sector}\label{sc:hnls}

In the so-called \emph{fermion portal}, new heavy neutral leptons (HNLs) are added to the SM to provide an elegant way to generate non-zero neutrino masses via the seesaw mechanism.~\cite{Mohapatra:1979ia,Yanagida:1980xy,Schechter:1980gr} In type-I seesaw models, one extends the SM by adding neutral right-handed fermions (identified with HNLs) that couple to the SM neutrinos similarly to the coupling between left- and right-handed components of the charged leptons. In a considerable class of these models, the HNLs become long lived and the LLP experiments have good prospects to detect them.~\cite{Beacham:2019nyx,Kling:2018wct,Hirsch:2020klk,Dercks:2018wum,Arbelaez:2019cmj,Cottin:2021lzz} For a recent review, the reader is referred to Ref.~\citenum{Cottin:2021lrq} and to references therein.

To highlight an example, in the case of right-handed neutrinos being produced in the decay of an additional $Z'$ boson in the gauged $B-L$ model, which also contains a singlet Higgs that spontaneously breaks the extra $U(1)_{B-L}$ gauge symmetry, \mbox{MAPP-2} will fill the gap left by CMS, LHCb, MATHUSLA, FASER2 and \mbox{CODEX-b}.~\cite{Deppisch:2019kvs} 

Light sterile neutrinos can account for dark matter,~\cite{Drewes:2013gca,Kusenko:2009up,Drewes:2016upu} while sterile neutrinos with a broad range of masses can account for the baryon asymmetry of the Universe through leptogenesis.~\cite{Deppisch:2015qwa} Sterile neutrinos may be long-lived in simplified models where the SM is extended with one sterile neutrino~\cite{Cottin:2018nms} or in neutrino-extended SM Effective Field Theories, $\nu$SMEFT.~\cite{Zhou:2021ylt} Intermediate-mass can be produced in leptonic and semi-leptonic decays of charmed and bottomed mesons, decaying to leptons via neutral and charged weak currents, thus becoming detectable in LHC LLP experiments.~\cite{DeVries:2020jbs} 

\subsection{Axions \& axion-like particles}\label{sc:alps}

Axions, or more generally ALPs, are pseudoscalar pseudo-Nambu-Goldstone bosons arising from approximate Abelian global symmetries beyond the SM which are broken spontaneously at a scale much greater than the electroweak scale. Axions, in particular, were postulated by the Peccei-Quinn theory to resolve the strong CP problem in quantum chromodynamics. ALPs also provide an interesting connection to the puzzle of dark matter, because they can mediate the interactions between the DM particle and SM states and allow for additional annihilation channels relevant for the thermal freeze-out of DM. Prospects for searches for ALPs in LLP experiments are reviewed in Refs.~\citenum{Alekhin:2015byh,Feng:2018pew,Aielli:2019ivi,Beacham:2019nyx,FASER:2019aik,Alimena:2019zri,Agrawal:2021dbo,Curtin:2018mvb}. 

\subsection{Supersymmetry}\label{sc:susy}

Going even further, supersymmetry (SUSY), a theoretical framework that provides a natural DM candidate, predicts the existence of LLPs. For instance, sleptons, charginos and R-hadrons (namely gluinos, top- or bottom-squarks) with unit electric charge may be detected in the MoEDAL NTDs, in the case they are sufficiently long lived to reach the detector.~\cite{Sakurai:2019bac,Felea:2020cvf,Acharya:2020uwc} 

$R$-parity violating SUSY also predicts LLPs, such as light long-lived neutralinos \none decaying via $\lambda'_{ijk}$ couplings to charged particles. Benchmark scenarios related to either charm or bottom mesons decaying into \none have been considered, in similar fashion as in sterile neutrinos, showing that these experiments can cover various meson production and decay modes and \none lifetimes~\cite{Aielli:2019ivi,Dercks:2018wum,Dercks:2018eua,Dreiner:2020qbi}.

\section{Post-discovery matters}\label{sc:discrim}

The variety of BSM scenarios predicting LLPs and the sensitivity to those by dedicated LLP experiments --- and main detectors equally --- begs the question regarding the power to provide qualitative and even quantitative information about the \emph{nature} of these LLPs in the post-discovery phase. The definition of \emph{LLP simplified models}~\cite{Alimena:2019zri} serves as a bridge between specific theoretical models and particular experimental signatures. 

The geometry of two-body $\gamma_{\rm d}/\phi$ decays to massless final states can provide information about their velocity and the ability to discriminate between different $\gamma_{\rm d}/\phi$ masses. CODEX-b can reconstruct the $\gamma_{\rm d}/\phi$ velocity to better than 1\% using spacial tracking information. If RPC timing information is used in a complementary way to discriminate between slow-moving new states, it is possible to separate between $\phi$ masses of 0.5~\gev and 2.0~\gev.~\cite{Gligorov:2017nwh}

If the LLP is pair-produced in Higgs boson decays, it is possible to measure the mass of this particle and determine the dominant decay mode with less than 100~observed events in MATHUSLA. In more general cases, the latter should be able to distinguish the production mode of the LLP and to determine its mass and spin based on the decay products of the long-lived particle.~\cite{Curtin:2017izq} 

Moreover, if information from the LLP and the main experiment of the same IP, e.g.\  MATHUSLA and CMS, is combined, the LLP production mode topology could be determined with as few as $\sim100$ observed LLP decays. Underlying theory parameters, like the LLP and parent particle masses, can also be measured with $\lesssim10\%$ precision.~\cite{Barron:2020kfo}

\section{Summary and Outlook}\label{sc:summary}

There is an ever increasing interest in long-lived particle searches at the LHC (and not only) to exlore the dark sector besides other unanswered questions in Particle Physics today. Besides the efforts in the main experiments ATLAS, CMS and LHCb, additional complementary experiments have been approved (MoEDAL, FASER, SND@LHC), aiming at data taking in Run~3, or have been proposed (MATHUSLA, CODEX-b, milliQan, ANUBIS, AL3X and others). In addition to these LHC detectors, the SHiP experiment at the SPS is also planned to explore hidden (dark) sectors. Several of these experiments have constructed, operated and analysed the data from small-scale demonstrators that proved the detector concept and provided the first encouraging (physics) results. Others plan to install such prototype for (part of) Run~3, before preparing a complete detector for the HL-LHC operation. The enigma of dark matter is challenged in a different perspective --- the lifetime frontier --- at the LHC Run~3 and beyond.  

\section*{Acknowledgements}

The author would like to thank the MG16 Meeting organisers for the kind invitation to present this talk. This work was supported in part by the Generalitat Valenciana via a special grant for MoEDAL and via the Project PROMETEO-II/2017/033, and by the Spanish MICIU / AEI and the European Union / FEDER via the grant PGC2018-094856-B-I00.


\bibliography{mitsou-lhc-llp}{}
\bibliographystyle{ws-mpla-vaso}

\end{document}